\magnification=\magstep1
\baselineskip=22pt
\parindent=0pt

\noindent
In Praise of Unstable Fixed Points: The Way Things Actually Work 
\bigskip
\noindent
Philip W. Anderson
\smallskip
\noindent
Henry Laboratories of Physics
\smallskip
\noindent
Princeton University, Princeton, NJ 08544
\bigskip
Abstract
\bigskip
\noindent
In recent years a fashion has grown up to ascribe great
importance to ``quantum
critical
points" at $T=0$, at the boundary between the basins of
attraction to the stable fixed
points
of ordered ground states. I argue that more physical significance
in connecting
microscopic
interactions with observed phenomena lies in the common
phenomenon of partially
ordered
``liquid" states at higher temperatures, unstable phases which
define the relevant
degrees of
freedom and may order in many different ways as the temperature
is further lowered.
\bigskip

\noindent
Corresponding author:\hfill\break
P.W. Anderson, Princeton University, Department of Physics,
Princeton, NJ 08544\hfill\break
Fax: 609-258-1006; email pwa@pupgg.princeton.edu

\vfill\eject

In our field of strongly-correlated electronic phenomena, the
first fad of the 21$^{st}$
century is the Quantum Critical Point.[1] (See Fig.~1.) This is
defined as a point (along a
line
representing different values of some control parameter) where
two ground states
with different
symmetries and different order parameters meet.  The argument is
that in the
neighborhood of
such a point we can show that the Ginzburg-Landau equations of
the system have a
higher
symmetry and also have the scale-invariance appropriate to such a
critical point.
Any unusual behavior such as unexpected power laws and absence of
conventional
excitation
spectra may be thus ascribed to the baleful influence of such a
critical point. The
``effective action"  (I don't know why such theories are always
Lagrangian rather than
Hamiltonian) which controls the low-energy behavior is restricted
to being a functional
of
the appropriate ``relevant variables": the order parameters of
the ordered phases.
The world,
then, is seen to be controlled ``from the bottom up": everything
is described in terms
of the
low-energy degrees of freedom of the ground states, as in Fig.~1.
The QCP is often
mysterious,
or hidden by the intrusion of yet another phase, but its
influence expands as $T$ is
raised
until its critical region encompasses the whole high $T$ phase
diagram.  All relevant
behavior
is controlled by some (D+Z) dimensional field theory; i.e., all
we need to do is find
what 
crank to turn on some universal dream machine.

As far as I know, there are no unequivocal examples of Fig.~1.
Critical regions near
$T=0$ are
normally quite narrow.[2] But perhaps more important is the
fundamental ontological
question: What
is the purpose of our scientific endeavor? If we abandon
description of physics in
terms of the
microscopic systems and forces--electrons, ions, Coulomb
interactions--then what is
the point
of our science? And this abandonment is precisely the goal of the
QCP fad.
The original purpose of the RNG in the study of critical behavior
was to explain
universality of
exponents and yet retain the connection to microscopic physics;
but the QCP sect
abandons the idea
of the RNG as based on fundamental microscopic physics, ignoring
the starting point
of the RNG
and dealing only with its end result. Microscopic variables are
seen as ``irrelevant"
and
``therefore" unimportant.

I would argue here that the physics of real condensed matter
systems is quite
otherwise. Fig.~1
is backwards. The spirit of the RNG is indeed correct: One wants
to start at high
energies and
temperatures, and integrate  out high frequency degrees of
freedom in stages in
order to build
low-temperature models with manageable complexity and, hopefully,
understandable
behavior. But
actual systems almost invariably fail to flow smoothly from high
energy to $T=0$.
There is a
strong tendency to hesitate in the neighborhood of an
infrared-unstable fixed point,
which is
hopefully described by some effective ``model" Hamiltonian which
still has a large
number of
relevant degrees of freedom. This model is often the result of a
projective
transformation of
the problem, removing certain high-energy degrees of freedom and
replacing them
with
constraints. But it cannot be correctly viewed as a functional of
the limited degrees of
freedom of an order parameter with a point symmetry, no matter
how complex.  As
we shall see,
such phases are liquids, paramagnetic insulators, rare earth
metals at normal
temperatures,
even the metallic Fermi liquid. I use the word ``phase" advisedly
because in many
cases we
recognize these as thermodynamically distinct phases--as e.g. the
liquid state, which
has no
symmetry distinction from the gaseous state but is physically
different.

So the world is more frequently described by a diagram like Fig.~2.  
One goes from
microscopic
physics via a projective transformation to an intermediate phase,
which is not ordered
but is
already a renormalized description satisfactory in some
intermediate range of
energies and
temperature--but is not a stable fixed point. At all but a few,
specially tuned values of
the
system parameters, this phase evolves further into one of several
different ordered
phases.

Let me illustrate this general structure in a number of cases of
decreasing simplicity
and
familiarity.  Fig.~3 shows the structure which is familiar to us
in thousands of materials:
molecular gases,  organic and polymer systems, and the like. 
Starting from electrons
and
nuclei, one eliminates electronic degrees of freedom via the
Born-Oppenheimer
approximation, 
arriving at
molecules with given intermolecular forces. (In relatively rare
cases, perhaps arriving
at a
covalently bonded network such as amorphous silicon or $S_i
O_2$.) Thus one is left
with no
electronic degrees of freedom at all.  In general, the attractive
parts of these forces
produce
a dense phase without crystalline order, which at lower
temperature orders, generally,
into
one of the hundreds of possible crystal structures.  The only
relevance to our further
reasoning is to observe how meaningless it is to think of the
liquid phase as  
arising out of quantum criticality at one or more of the
(first-order) phase transitions
between different crystal structures.

Fig.~4 illustrates the next simpler case: the Fermi liquid.
Here the electronic degrees of freedom are not all gapped, so one
has to eliminate
the 
high-frequency ones via the Galivotti-Shankar
renormalization[3] and
one is left with the
Landau
Fermi liquid Hamiltonian plus quasiparticle scattering, for
quasiparticles which are
the
residual degrees of freedom. (There are also, of course, lattice
degrees of
freedom--the
phonons. Fortunately,  Migdal's theorem, shows that these can
be treated
perturbatively.)

But the Fermi liquid in general is not stable. It may further
renormalize into not only
$L=0$
BCS but any of the angular-dependent BCS states, depending on the
nature of the
residual
interaction. It may also have SDW or ferromagnetic instabilities.
Again, I would
challenge any
one to reconstruct the Fermi liquid as resulting from a QCP
between any of these
phases.

Now we come to a more complex case: ``Mixed valence" (see Fig.~5). 
We start with
actinide 
or lanthanide
ions and metal electrons. We bypass the complications of the true
``mixed valence"
state, which
renormalizes to a relatively simple metal (as shown by Haldane). 
That is, in Fig.~5 renormalization paths near the edge of the
Anderson Model regime
can flow
into the ``Kondo valence collapse" regime. But for reasonably
symmetric systems 
renormalization is via a Schrieffer-Wolff transformation[4] to the
``Kondo lattice." 
This unstable fixed point is massively non-Fermi liquid: the
lanthanide or actinide f 
electrons are represented
by pure spins with no charge degrees of freedom, these latter
having been projected
away by the
Anderson $\to$ Kondo renormalization.

But now we find that in many cases the Kondo Hamiltonian
renormalizes again via the
$T_K$
mechanism to a new Fermi liquid!  On the other hand, it may also
happen that
residual spin-spin
interactions enforce ferromagnetism or antiferromagnetism. In a
quite separate way,
the Kondo
$FL$ is itself unstable to non BCS SC or to SDW or both. (Note
that BCS does not
happen: the
spin-derived electrons do not interact strongly with phonons.)

Of course, I have been leading up to still another unstable fixed
point: the RVB phase:
a
lightly-doped 2D Mott insulator. In this case the starting
high-temperature Hamiltonian
has
strong local repulsion--a Hubbard model, with $U >> t$--and the
appropriate initial
renormalization is the $t-J$ Anderson-Rice
transformation,[5] which
leaves us with a
truncated
Hilbert space, containing hole and spin degrees of freedom and a
$J$ interaction
beteen spins. 
As noted in a recent article, for
low density the resulting phase renormalizes to a Fermi liquid,
or, in 2D, to a Fermi
liquid-like phase. (For low density, the Galitskii-Lee-Yang
pseudopotential  
scheme is correct.)
But for high density, a spin pseudogap phase may develop, which
is distinct from the
Fermi
liquid and can condense into several different ordered phases:
N\'eel
antiferromagnet, d-wave
superconductor, inhomogeneous stripe phases, and possibly the
staggered flux
phase of Lee and
Nayak et al.[6]  The pseudogap phase is best described, probably, as
having
d-symmetry
condensation of spinon pairs.  It is not ``caused" in any sense
either by the SC-AF
quantum
critical point nor by the crossover to a true Fermi liquid(like)
state.  Fig.~6 is a sketch
of
the appropriate phase relationships.

In conclusion, I would like to remark that none of the ``liquid"
phases (except possibly
the
Fermi liquid) is easy to analyze in any precise way--this has
long been known for the
true 
liquid, for instance, and we are only just learning it for the
Mott paramagnet.  This
however
does not imply that they have no identity or physical reality. 
The simplicity of BCS
comes
from the simplicity of the Fermi liquid state from which it
arises; the difficulty of the
calculation of melting points, in contrast, comes from the
relatively mysterious nature
of the
liquid state. So we cannot expect high $T_c$ cuprates to be
simple; but we find that
their
physical motivation is not mysterious. 

\vfill\eject

References
\bigskip
\item{[1]} See for instance Subir Sachdev,  "Quantum Phase Transitions", 
Cambridge University Press,  Cambridge, 1999.
\medskip

\item{[2]} D. Huse, private communication.
\medskip

\item{[3]} G. Benfatto and G. Gallivotti, J. Statist. Phys 5{\bf 9}, 541(1990); 
R.  Shankar, Physica A {\bf 177}, 530 (1991)
\medskip

\item{[4]} J.R. Schrieffer and P.A. Wolff, Phys. Rev. {\bf 149}, 491 (1966)

\item{[5]} C Gros, R. Joynt and T.M. Rice, Phys. Rev. {\bf 381} (1987); see
also J.E. Hirsch, Phys. Rev. Lett. {\bf 54}, 1317 (1985),  and G Baskaran Z Zou and
P.W. Anderson, Sol. St. Comm. {\bf 63}, 973 (1987).
\medskip

\item{[6]} P.A. Lee and X.-G. Wen, Phys. Rev. Lett. {\bf 76},503 (1996)
\vfill\eject

Figure captions

\item{Fig. 1} The Quantum Critical Point Picture. Above the QCP at $T=0$ is an
Ever-expanding
``Critical
Region" of Anomalous Properties.

\item{Fig. 2} The Intermediate Unstable Fixed Point Picture. Electrons and Ions
Renormalize to an UFP, or
Model, with Constrained Dynamics but $\approx N$ Degrees of
Freedom, Which
Then Further
Condenses to Various Ordered States.

\item{Fig. 3} The Molecular Liquid State as an Example of the UFP: The
Born-Oppenheimer
Projection Leaves Us
with Molecular Motions Only. Molecular Liquid Can Condense into
Many Phases.

\item{Fig. 4} The Fermi Liquid as Example: Gallivotti-Shankar Projection Leaves
Us with Quasiparticles Near a
Fermi Surface, Which Then Can Condense into Ferromagnetism, Spin-
or
Charge-Density Waves or
BCS with Any L.

\item{Fig. 5}
The Kondo Hamiltonian as an UFP. This Hamiltonian Has f-Electric
Charge
Fluctuations Completely
Projected Out, Yet Can Renormalize to an Unusual Fermi Liquid,
SDW, or $L \not =
0$ BCS
Superconductor.

\item{Fig. 6} The RVB-Pseudogap State as an UFP, Only in the Low-Doping Regime
of the
1-Band Hubbard Model.
It Can Condense to Stripes, d-Wave Superconductor or N\'eel
Antiferromagnet.

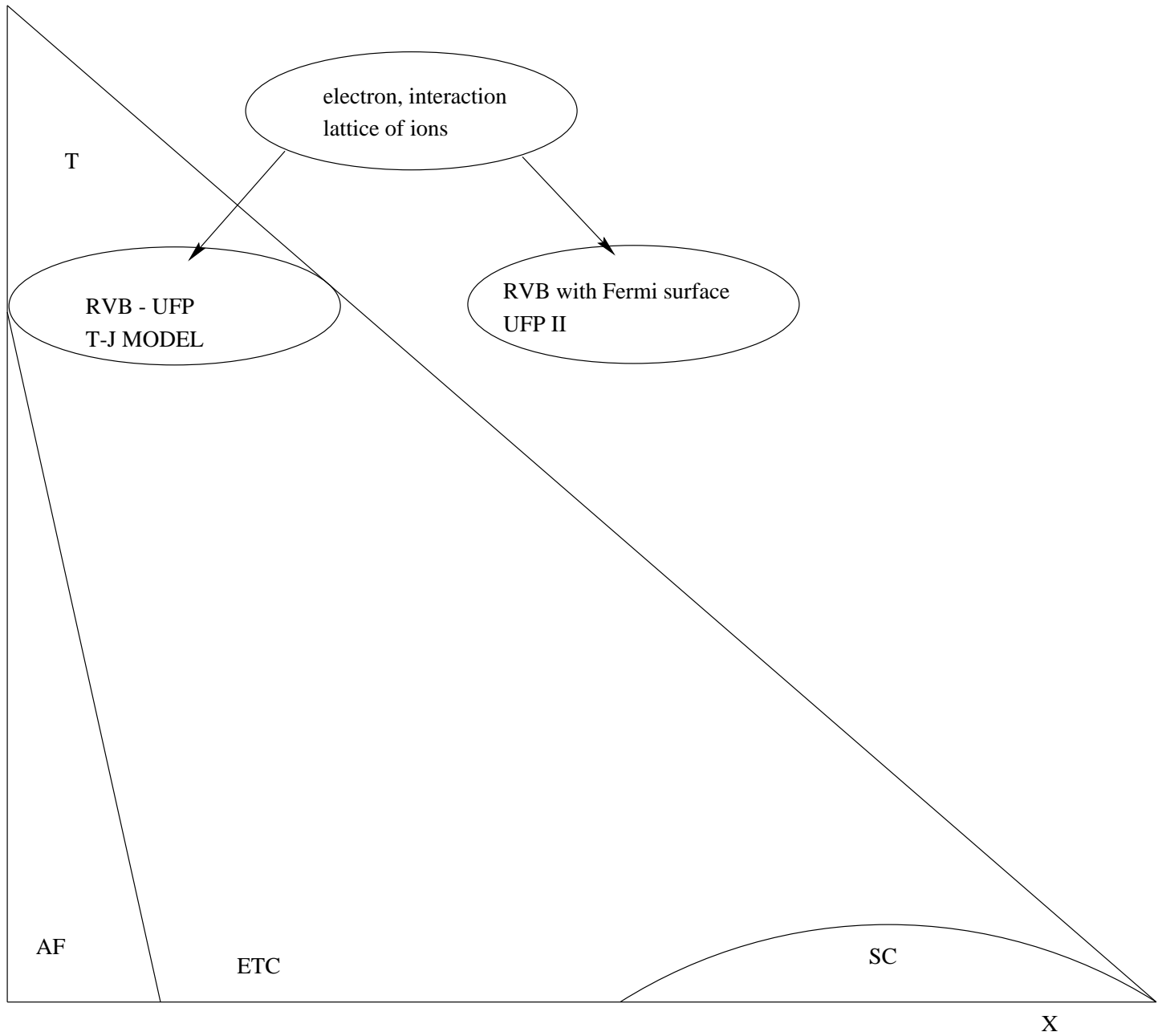
\end{document}